\documentclass[a4paper]{article}

\usepackage{multirow}
\usepackage[dvipsnames]{xcolor}
\usepackage{INTERSPEECH2020}
\usepackage[printonlyused,withpage]{acronym}
\usepackage{soul}

\title{Exploiting Fully Convolutional Network and Visualization Techniques on Spontaneous Speech for Dementia Detection}
\name{Youxiang Zhu, Xiaohui Liang}
\address{
  Department of Computer Science, University of Massachusetts Boston, USA
\email{\{Youxiang.Zhu001, Xiaohui.Liang\}@umb.edu}
}

\begin{document}

\maketitle
 
\begin{abstract}
In this paper, we exploit a \ac{FCN} to analyze the audio data of spontaneous speech for dementia detection. A fully convolutional network accommodates speech samples with varying lengths, thus enabling us to analyze the speech sample without manual segmentation. Specifically, we first obtain the \ac{MFCC} feature map from each participant's audio data and convert the speech classification task on audio data to an image classification task on \ac{MFCC} feature maps. Then, to solve the data insufficiency problem, we apply transfer learning by adopting a pre-trained backbone \ac{CNN} model from the MobileNet architecture and the ImageNet dataset. We further build a convolutional layer to produce a heatmap using Otsu's method for visualization, enabling us to understand the impact of the time-series audio segments on the classification results. We demonstrate that our classification model achieves 66.7\% over the testing dataset, $>62.5\%$ of the baseline model provided in the ADReSS challenge. Through the visualization technique, we can evaluate the impact of audio segments, such as filled pauses from the participants and repeated questions from the investigator, on the classification results.

\end{abstract}

\acresetall

\section{Introduction}\label{sec:introduction}

The number of patients with \ac{AD} over the age of 65 is expected to reach 13.8 million by 2050, causing a huge crisis on the public health system~\cite{2020AD}. While there is no proven effective treatment on \ac{AD}, it is significant to detect early symptoms of \ac{AD} such that interventions can be implemented in the early stage. While screening measures, neuropsychological assessments, and MRI imaging scans are not pragmatic approaches, recent studies have explored spontaneous speech for a practical and low-cost early detection of dementia symptoms. Pitt corpus~\cite{pitt}, one of the large speech datasets, includes spontaneous speech obtained from a \ac{CTP} description task. The \ac{CTP} task has also been explored with computerized agent to automate and mobilize the speech collection~\cite{mirheidari2017avatar,mirheidari2019detecting} and in other languages including Mandarin~\cite{wang2019towards,chien2019automatic}, German~\cite{ILSE}, Sweden~\cite{fraser2019multilingual}. Other spontaneous speech datasets for AD research include those collected from film-recall tasks~\cite{toth2018speech}, story-retelling tasks~\cite{fraser2013automatic}, map-based tasks~\cite{de2019protocol}, and human conversations~\cite{mirheidari2019dementia}. 

Researchers have studied linguistic features extracted from transcripts for building classification and regression models. A recent survey showed effective linguistic features include semantic content, syntax and morphology, pragmatic language, discourse fluency, speech rate, and speech monitoring~\cite{mueller2018connected}. The linguistic features were often manually selected based on expert knowledge, and the analysis methods were complex and highly task-dependent. A potential research direction is to automate the linguistic analysis. For example, Croisile et al. manually extracted 23 information units from the picture using language knowledge and found the analysis based on them was effective in dementia detection~\cite{croisile1996comparative}. Fraser et al. confirmed both the initial 23 information units and the auto-generated information units are effective in analysis~\cite{fraser2019embeddings}. Yancheva et al.~\cite{yancheva2016vector} and Fraser et al.~\cite{fraser2019multilingual} further proposed to auto-generate topic models that can recall 97\% of the human-annotated information units. Similarly, the acoustic-based analysis was started with pre-defined features and recently automated with computational models. Hoffman et al. considered acoustic features for each utterance~\cite{hoffmann2010temporal}. Fraser et al. evaluated the statistical significance of pause and word acoustic features~\cite{fraser2013automatic}. Toth et al. considered four descriptors for silent/filled pauses and phonemes~\cite{toth2015automatic}. Toth et al. implemented a customized automatic speech recognition (ASR) and automatic feature selection for phones, boundaries, and filled pauses~\cite{gosztolya2016detecting,toth2018speech}. Haider et al. proposed an automatic acoustic analysis approach using the paralinguistic acoustic features of audio segments~\cite{haider2019assessment,luz2020alzheimers}. 

In this paper, we envision an automated speech analysis of the audio data for dementia detection. We observed Haider et al. segmented the audio data into small pieces by setting the log energy threshold parameter to 65dB with a maximum duration of 10 seconds~\cite{haider2019assessment,luz2020alzheimers}. We feel the segmentation may cause critical time-series information loss. Any smaller speech segments hardly represent the overall speech sample. In addition, the speech continuity is removed by segmentation, making the model inaccurately capture the time-series characteristics. Thus, our model aims to accommodate a speech sample of each participant as input and preserve the time-series characteristics of the speech samples~\cite{hershey2017cnn,zhang2018attention}. Our contributions are as follows.

\begin{figure*}[!htbp]
    \centering
    \includegraphics[width=0.85\textwidth]{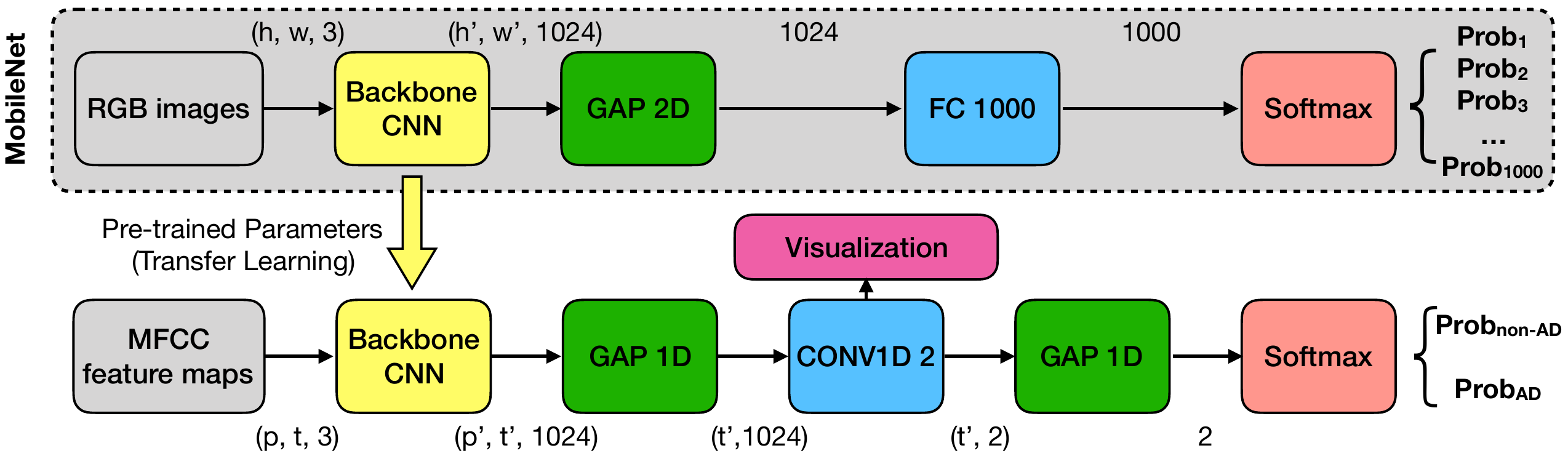}
    \caption{Proposed classification model with transfer learning and visualization}
    \label{fig:model}
        \vspace{-0.4cm}
\end{figure*}

First, we converted a speech classification task on the audio data to an image classification task on the \ac{MFCC} feature maps. The feature maps are automatically extracted from the audio data and preserve the time-series characteristics of the speech.

Second, we explored the \ac{FCN} to accommodate the speech samples with varying lengths. We employed the transfer learning technique by adopting a pre-trained backbone \ac{CNN} from the MobileNet architecture and the ImageNet dataset. Compared to the baseline model, ours achieves better accuracy and a more balanced F1 score.

Third, we embedded a convolutional layer in our model to enable the visualization of the impact of audio segments on the classification results, thus increasing our understanding of how the classification model works. We found the visualization technique identifies the filled pauses from the participant and the repeated questions from the investigator as positive signs of \ac{AD}.
\section{ADReSS Challenge Dataset}\label{sec:dataset}

We studied the dataset created for the ADReSS challenge~\cite{luz2020alzheimers}, which is a part of the Pitt corpus~\cite{pitt}, with the numbers of participants balanced for age and gender. The data consists of speech recordings and transcripts of spoken picture descriptions elicited from participants through the Cookie Theft picture from the Boston Diagnostic Aphasia Exam~\cite{honig2001natural,goodglass2001bdae}. We studied the full-wave enhanced audio, which contains the audio recordings after noise removal. The training dataset includes speech data from 24 male participants with AD, 30 female with AD, 24 male non-AD participants, and 30 female non-AD participants. The ADReSS testing dataset includes speech data from 11 male participants with AD, 13 female with AD, 11 male non-AD participants, and 13 female non-AD participants. The complete dataset information can be found at Luz et al.~\cite{luz2020alzheimers}.
\section{MFCC Feature Maps}\label{sec:feature}

\begin{figure}[!htbp]
    \centering
    \includegraphics[width=0.4\textwidth]{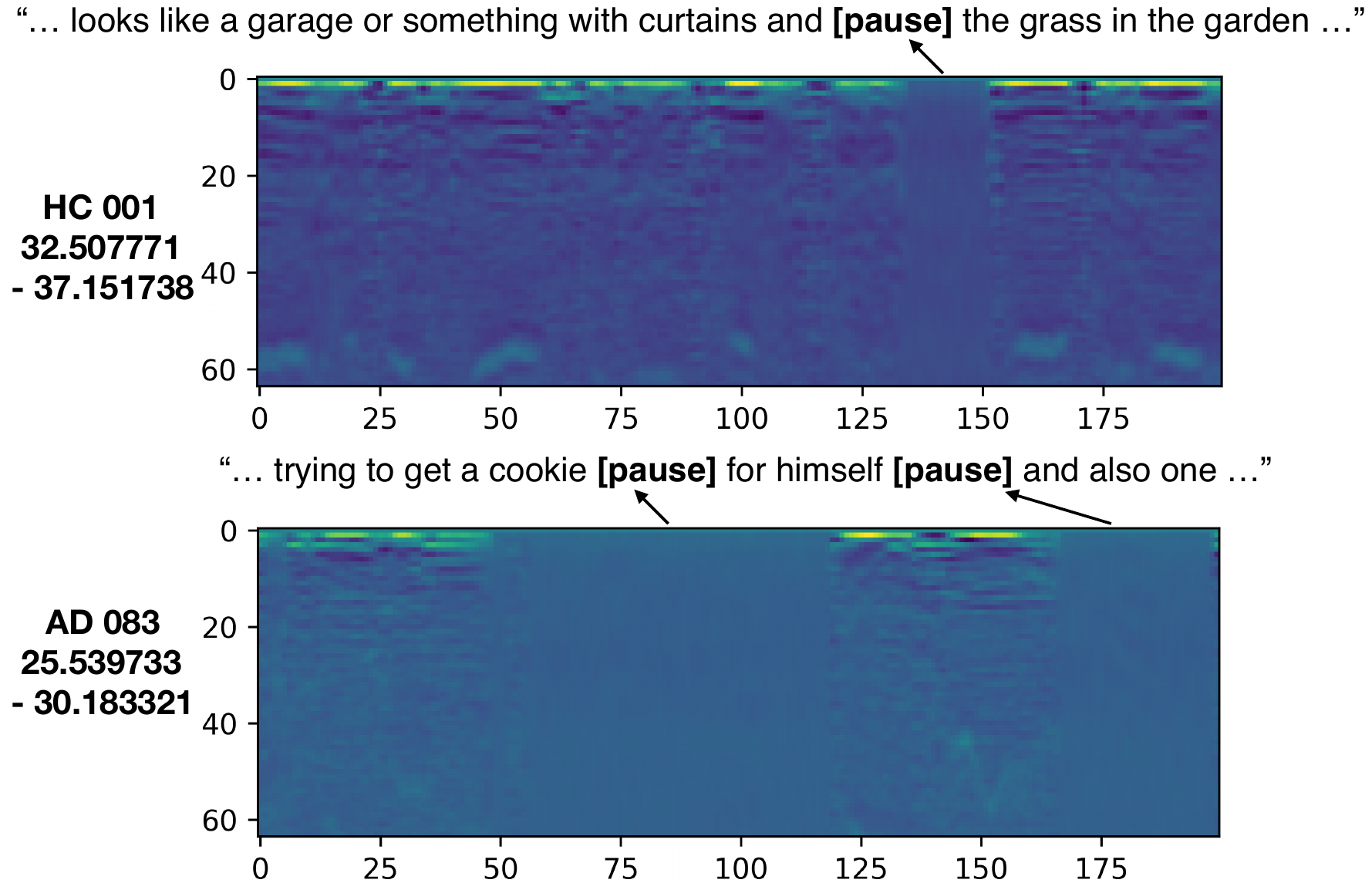}
    \caption{Sample MFCC feature maps of 001 and 083}
    \label{fig:feature}
    \vspace{-0.5cm}
\end{figure}

Mel-frequency cepstral coefficients have been widely used in speech recognition research~\cite{muda2010voice}. Fraser et al. carried out an acoustic-prosodic analysis on the Pitt corpus using 42 MFCC features~\cite{yancheva2016vector,fraser2016linguistic}. We extracted an MFCC feature map for each participant's entire audio sample. The MFCC feature map is denoted as a $(p,t)$-matrix where the hyper-parameter $p$ is set to 64, and $t$ is related to the duration of the speech sample. We use librosa function with sampling rate of 22050, window size of 2048, and step size of 512. In Figure~\ref{fig:feature}, we show the sample \ac{MFCC} feature maps of participants 001 (non-AD) and 083 (AD), respectively. The data in the first row is scaled for visualization purposes. By extracting the MFCC feature maps, we convert the speech dataset to an image dataset. The advantages of \ac{MFCC} feature maps are three-fold: i) the conversion from speech to \ac{MFCC} feature maps can be done automatically; ii) the silent pauses in the audio data are preserved as a distinctive feature in \ac{MFCC} feature maps as shown in Figure~\ref{fig:feature}; iii) we found the audio dataset contains speech from the investigator and filled pauses from the participant that are shown to be important~\cite{toth2018speech}. While identifying these audio segments requires expensive human efforts or customized \ac{ASR}, we envision the \ac{MFCC} feature maps preserve the time-series structure, and the classification model may continuously learn to deal with these effects.

\section{Classification Model}\label{sec:model}

We aim to design a classification model to classify the audio samples into the non-AD and AD groups. After converting each audio sample to an MFCC feature map, we focus on developing an image classification model. To improve the learning effectiveness over the small audio dataset, we apply transfer learning using ImageNet and MobileNet. In the following, we first explain the transfer learning technique and then introduce our model. An overview of our model is shown in Figure~\ref{fig:model}.

\subsection{Transfer learning via ImageNet and MobileNet}

We developed a transfer learning technique using the knowledge from the \textit{image datasets} and \textit{pre-trained image classification models} to overcome the insufficiency of the audio dataset. 

\textit{ImageNet} is an image dataset organized according to the WordNet hierarchy~\cite{imagenet_cvpr09}. Each meaningful concept in WordNet, possibly described by multiple words or word phrases, is called a ``synset.'' There are more than 100,000 synsets in WordNet, majority of which are nouns (80,000+). ImageNet provides, on average, 1000 images to illustrate each synset. Images of each concept are quality-controlled and human-annotated. ImageNet dataset has been widely used in designing and evaluating the image classification models~\cite{ILSVRC15}. \textit{MobileNet} is a lightweight network architecture that significantly reduces the computational overhead as well as parameter size by replacing the standard convolution filters to the depth-wise convolutional filters and the point-wise convolutional filters~\cite{howard2017mobilenets}. The total parameters of the MobileNet backbone are of a size 17.2 MB, significantly less than other convolutional neural networks. Considering the limited size of the speech dataset, we thought a smaller model with less complexity, such as MobileNet, may worth being tested.

The MobileNet architecture is shown at the above layer of the Figure~\ref{fig:model}. With an RGB image as input, the output is the probability that the image belongs to each of the 1000 classes. Denote the input image as a 3-dimensional $(h, w, 3)$-matrix where $h$ is height, $w$ is width, and 3 represents the RGB channel. A backbone \ac{CNN} consists of a set of convolution, pooling, and activation operations. We used the full width (1.0) MobileNet backbone pre-trained on a resolution of 128*128 images. The detailed architecture can be refer to the paper~\cite{howard2017mobilenets}. The backbone converts an input of $(h, w, 3)$-matrix to an output of $(h',w',1024)$-matrix where $(h',w')$ are functionally related to $(h,w)$, and 1024 represents the feature channel number, i.e., the depth of the backbone \ac{CNN}. The output $(h',w',1024)$-matrix is then fed to a \ac{GAP} layer for reducing the dimensions of $h'$ and $w'$ and obtaining a $1024$-dimension feature. A \ac{FC} layer with 1000 neurons is employed to produce the output according to the wanted 1000 classes. Lastly, a softmax activation layer is added to produce the classification results as the probabilities for 1000 classes that add up to 1. The pre-training of MobileNet is time-consuming and may take weeks due to the large ImageNet dataset. The pre-trained parameters of the backbone \ac{CNN} from MobileNet are made available, though. We used the parameters and saved time on the pre-training.

\subsection{Proposed model}

Our proposed model is shown at the bottom layer of the Figure~\ref{fig:model}. Our FCN architecture employs the pre-trained backbone CNN module from the MobileNet. Denote the \ac{MFCC} feature map of the audio sample as a $(p, t, 1)$-matrix where $p$ is a hyper-parameter set to 64, and $t$ is related to the duration of the speech sample. To match with the module input, i.e., an RGB image, we duplicated
the \ac{MFCC} feature map twice and made the \ac{MFCC} feature map as a $(p,t,3)$-matrix. In this way, we can feed the \ac{MFCC} feature map into the backbone CNN module of the MobileNet in the same way as an RGB image. The output of the backbone \ac{CNN} is denoted as a $(p',t',1024)$-matrix where $(p',t')$ are functionally related to $(p,t)$. We employed a \ac{GAP}-1D (one-dimensional) to reduce $p'$ dimension of the matrix. The $t'$ dimension is preserved for enabling the visualization. We further used a 1D convolutional layer with $2$ neurons to adapt to the wanted 2 classes. The output of the 1D convolutional layer is used to build a 1D heatmap for visualization. Finally, we added another GAP-1D layer to reduce the $t'$ dimension and the softmax activation layer to produce the classification results as two probabilities for the two classes that add up to 1.
\section{Evaluation}\label{sec:evaluation}

\begin{figure*}[!htbp]
    \centering
    \includegraphics[width=0.95\textwidth]{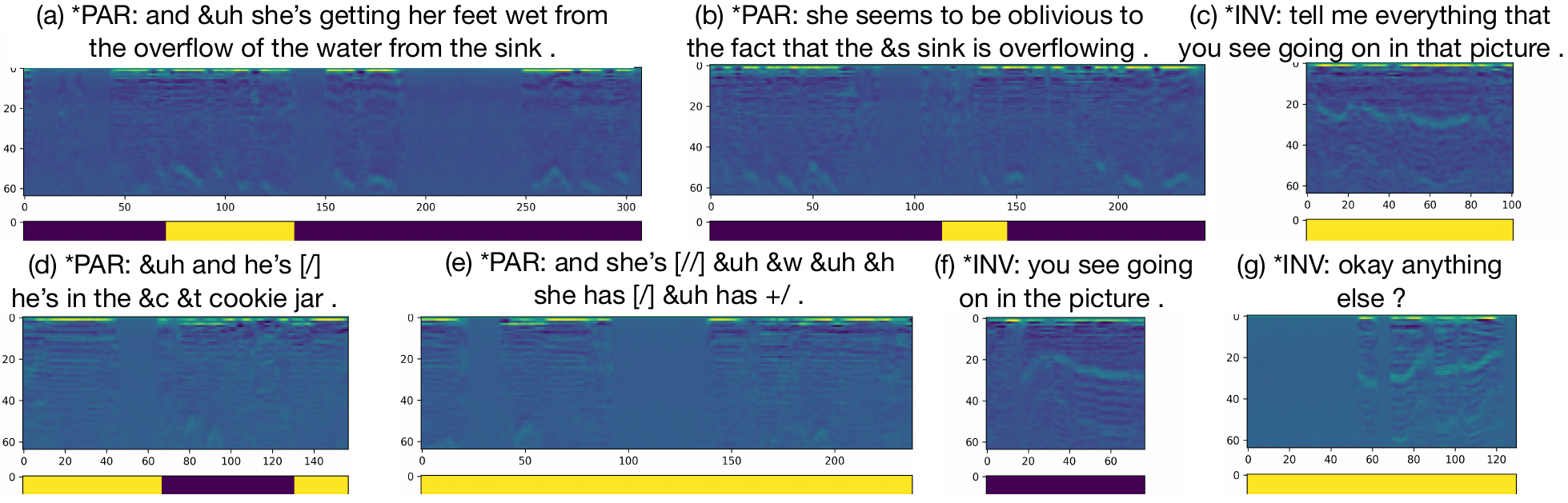}
    \caption{Visualizing the impacts of audio segments. (a,b,c) are from 001 non-AD and (d,e,f,g) are from 079 AD}
    \label{fig:vis}
    \vspace{-0.5cm}
\end{figure*}

\subsection{Implementation and training strategy}

We implemented the classification model with Keras and TensorFlow. We used a mini-batch with batch size $8$ at each training step and a very small learning rate of 1e-5, while minimizing the cross-entropy loss with the RMSProp optimizer~\cite{tieleman2012lecture}. As the samples in a mini-batch are required to have the same size, we used zero-paddings to pad the samples such that their lengths are equal to the max length in the mini-batch. The zero-paddings have limited impact on the classification task because i) they can be easily distinguished from non-zero pauses and ii) the employed GAP and softmax layers produces average height and width values and relative values. We further confirmed from our visualization techniques that zero-paddings in the training produced minimum impact on the classification results. In the testing phase, we treated the testing sample as a mini-batch with a batch size of $1$ such that our classification model can take samples with any length.

Due to the limited audio dataset, we designed a set of training strategies as follows. First, we split the provided speech dataset into two halves with equal sizes. We used one half for training and the other half for validation. We trained our model with max 1000 epochs and selected the epoch that has the highest validation accuracy after the model converges. We performed such training twice by switching the training and validation datasets. We thus obtained two models $(M_1, M_2)$ that are a complement to each other. Note that $k$-fold cross-validation is a classical evaluation strategy. Here, we chose $k=2$ for a relatively large validation dataset because a too small validation dataset may not reflect the overall data distribution. A larger validation dataset enables us to better control the learning rate for our model to produce a stable classification accuracy on the validation dataset. We did not use ``leave-one-subject-out (LOSO)" because i) LOSO is not suitable for training deep neural network models in terms of computational efficiency; ii) we focus on learning the parameters of the deep neural networks but not model selection; and iii) we focus on the results of the provided test dataset. Second, we merged the above two models $(M_1, M_2)$ into $M_{1+2}$ by averaging the output probabilities of $(M_1,M_2)$. $M_{1+2}$ takes an advantage of the entire dataset. Third, we used all the speech samples to train a model $M_3$, where we selected the epoch with minimum training loss instead of maximum validation accuracy. Lastly, we merged the above three models $(M_1, M_2, M_3)$ to a model $M_{1+2+3}$ by adding the output probabilities of $(M_1, M_2, M_3)$. Note that, models $(M_1, M_2, M_{1+2})$ mainly focus on improving the validation accuracy. This strategy is usually adopted when the training dataset is small. Models $(M_3, M_{1+2+3})$ consider minimizing the training loss, which is a general approach for deep learning, where a large training dataset is available.

\subsection{Classification results}

We trained and tested our models using two different datasets, both of which are provided by the ADReSS challenge. The testing dataset were provided after the models were trained. Our models output a binary result, non-AD or AD. The evaluation metrics are accuracy $\frac{TN+TP}{N}$, precision $\pi=\frac{TP}{TP+FP}$, recall $\rho=\frac{TP}{TP+FN}$, and F1 score $\frac{2\pi\rho}{\pi+\rho}$, where $N$ is the number of participants, $TP$, $FP$ and $FN$ are the numbers of true positives, false
positives and false negatives, respectively.

\vspace{-0.5cm}

\begin{center}
\begin{small}
\begin{table}[!htbp]
\caption{Classification results}
\vspace{-0.3cm}
\begin{tabular}{llllll} 
 \hline
  & Class & Prec. & Recall & F1 & Acc. \\
  \hline
 \multirow{2}{5em}{$M_1$ (val.)} & non-AD & 0.68 & 0.54 & 0.60 & \multirow{2}{2em}{\colorbox{Green}{0.646}} \\
  & AD & 0.62 & 0.75 & 0.68 & \\
  \hline
 \multirow{2}{5em}{$M_2$ (val.)} & non-AD & 0.58 & 0.58 & 0.58 & \multirow{2}{2em}{\colorbox{BurntOrange}{0.583}} \\
  & AD & 0.58 & 0.58 & 0.58 & \\
  \hline
 \multirow{2}{5em}{$M_{1+2}$} & non-AD & 0.63 & 0.79 & 0.70 & \multirow{2}{2em}{\colorbox{Green}{0.667}} \\
  & AD & 0.72 & 0.54 & 0.62 & \\
  \hline
 \multirow{2}{5em}{$M_3$ (loss)} & non-AD & 0.63 & 0.71 & 0.67 & \multirow{2}{2em}{\colorbox{Green}{0.646}} \\
  & AD & 0.67 & 0.58 & 0.62 & \\
  \hline
 \multirow{2}{4em}{$M_{1+2+3}$} & non-AD & 0.59 & 0.67 & 0.63 & \multirow{2}{2em}{\colorbox{BurntOrange}{0.604}} \\
  & AD & 0.62 & 0.54 & 0.58 & \\
  \hline
 \multirow{2}{5em}{Baseline~\cite{luz2020alzheimers}} & non-AD & 0.67 & 0.50 & 0.57 & \multirow{2}{2em}{0.625} \\
  & AD & 0.60 & 0.75 & 0.67 & \\
 \hline
\end{tabular}
\label{table:results}
\end{table}
\end{small}
\end{center}

\vspace{-0.8cm}

Our models $(M_1, M_2)$ achieve 64.6\%, 58.3\% accuracy, respectively, as shown in Table~\ref{table:results}. We found these are consistent with our validation accuracy 62.96\% and 61.11\% obtained in the training phase. After combining the outputs from $M_1$ and $M_2$, our model $M_{1+2}$ achieves the highest accuracy 66.67\% of our five attempts. We consider the model $M_{1+2}$ is relatively successful as it outperformed $M_1$, $M_2$, and the baseline model with 62.5\%. We think this performance gain of $M_{1+2}$ is obtained because it considers all samples in training and inherits the knowledge from the image classification model via transfer learning. Our models $M_3$ and $M_{1+2+3}$ achieves 64.6\% and 60.4\% accuracy, respectively. Without the validation step in training, these models focus on minimizing the training loss and definitely need more data to improve accuracy. Both models $M_{1+2}$ and $M_3$ used all the samples in the training phase, but $M_{1+2}$ achieves a higher accuracy than $M_{3}$. One possible explanation is that in $M_{1+2}$ after splitting the training samples into two equal halves, the sample-wise differences in each half become smaller. When $M_{1+2}$ merges the outputs of $M_1$ and $M_2$, it simply chooses the model with higher confidence and thus produces a higher accuracy. In general, as the training samples are limited in size and with a large sample-wise difference, our model may largely modify the parameters of the pre-trained backbone \ac{CNN} from the MobileNet, resulting in a degeneration of the discriminative ability of the pre-trained model and yielding overfitting. An enhanced data splitting method may help. Note that we currently split the training dataset into two 
equal-sized halves in a random way, and we envision that the \ac{CNN} feature-based 
splitting method may enhance the performance. At last, we found that our five models achieve more balanced F1 scores, compared to the baseline model~\cite{luz2020alzheimers}.
\section{Visualization}\label{sec:visualization}

One significant contribution of our models is to enable the visualization of the impacts of the audio segments on the classification results. As shown in Figure~\ref{fig:model}, our model incorporates a 1D convolutional layer with 2 neurons that converts a $(t',1024)$-matrix to a $(t',2)$-matrix, where $t'$ is functionally related to the time $t$. For the dimension of size 2, the first row represents the non-AD class, and the second row represents the AD class. We chose the second row (or the first row) and used the Otsu's thresholding method to evaluate the impact scores over the time dimension. Otsu's method is used to perform an automatic image thresholding~\cite{otsu1979threshold}. In the simplest form, the algorithm returns a single intensity threshold that separates pixels into two classes, foreground and background. This threshold is determined by minimizing intra-class intensity variance, or equivalently, by maximizing inter-class variance. In our visualization module, we used Otsu's method to assign either $0$ for smaller values (dark color) or $1$ for larger values (yellow color). We also used the nearest neighbor interpolation technique to scale the vector from size $t'$ to size $t$.

Figure \ref{fig:vis} shows the visualization bar on seven utterances. In general, for non-AD samples, the dark segments contribute more to the non-AD result compared to the yellow segments, for AD samples, the yellow segments contribute more to the AD result compared to the dark segments. Note that the Otsu's method produces both dark and the yellow segments for both non-AD and AD samples. We have two observations. First, from (a) and (b), dark segments represent quality speech and contribute to the non-AD results; from (d) and (e), yellow segments represent filled pauses and unclear speech and contribute to the AD results. Second, we have an interesting observation on the investigator's audio data, which are mixed with the participant's audio data. From (c), the investigator's speech mixed with the non-AD's sample shows yellow, contributing to the AD result as a noise. From (f), the investigator's speech mixed with AD's sample shows dark, contributing to the non-AD result as a noise. More importantly, from (g), the investigator's speech from the AD sample shows yellow, contributing to the AD result. By 
cross-checking the transcripts, we found our model may capture the similar utterance that the investigators used for pushing the AD participants for more conversations and consider this utterance as a positive sign for the AD result.
\section{Discussion}\label{sec:discussion}

\textbf{Data augmentation.} As the training samples are limited, we exploit possible data augmentation techniques. Common image augmentation techniques include rotating or zooming in/out the images for additional samples. However, these techniques do not apply to the \ac{MFCC} feature maps due to the different meanings of their different dimensions. Thus, we consider another data augmentation technique, i.e., randomly masking certain periods of an \ac{MFCC} feature map with zeros. The mask is randomly generated at different positions for every epoch with a length of 200 to 400 units. One advantage of using this data augmentation method is its consistency with the data representation of our model input, where zero-paddings were adopted in the mini-batch implementation. However, our attempts did not introduce significant accuracy gains.

\textbf{Heterogeneous speech dataset.} The limited speech dataset is the biggest barrier in this research. Researchers proposed different models and tested them over different datasets, resulting in loosely-connected conclusions. ADReSS challenge is a significant effort to bring researchers together to study the same dataset for producing more meaningful results. Our computational model is fully automated and has the potential to apply to any speech and even multilingual speech. While researchers explicitly identified silent pauses, filled pauses, and speech duration for building classification models, we envision these features are preserved in the feature maps, and as more datasets are available, our computational model will self-adapt to both explicit and implicit acoustic features.
\section{Conclusions}\label{sec:conclusion}

We proposed a classification model to analyze audio data for dementia detection. Our model employs the fully convolutional network to accommodate the audio samples with varying lengths and preserve the time-series characteristics. We extracted the MFCC feature maps from the audio data and converted the speech classification task to an image classification task. We then applied the transfer learning technique to adopt a pre-trained model from the MobileNet architecture. Our model achieves higher accuracy than the baseline model. Finally, we implemented a visualization technique to provide intuitive visual feedback on the impacts of the audio segments on the classification results. We envision our computational model can be applied to other speech datasets and have the potential to be continuously enhanced with deep learning techniques.
\section{Acknowledgements}

This research is funded by the US National Institutes of Health National Institute on Aging, under grant No. 1R01AG067416.

\bibliographystyle{IEEEtran}
\bibliography{main}

\newpage

\section{List of Acronyms}
\begin{acronym}
\acro{ADRD}{Alzheimer's Disease and Related Dementias}
\acro{AD}{Alzheimer's Disease}
\acro{MCI}{Mild Cognitive Impairment}
\acro{HC}{Health Control}
\acro{WLS}{Wisconsin Longitudinal Study}
\acro{CTP}{Cookie Theft Picture}
\acro{IVA}{Intelligent Virtual Agent}
\acro{IU}{Information Units}
\acro{MFCC}{Mel Frequency Cepstral Coefficient}
\acro{MFC}{Mel-Frequency Cepstrum}
\acro{LLDs}{Low-Level Descriptors}
\acro{LSP}{Line Spectral Pair}
\acro{AOI}{Area of Interest}
\acro{ASR}{Automatic Speech Recognition}
\acro{ML}{Machine Learning}
\acro{MMSE}{Mini-Mental State Examination}
\acro{MoCA}{Montreal Cognitive Assessment}
\acro{GDS}{Geriatric Depression Scale}
\acro{GAI}{Geriatric Anxiety Inventory}
\acro{SVM}{Support Vector Machine}
\acro{PCA}{Principal Component Analysis}
\acro{DNN}{Deep Neural Network}
\acro{FCN}{Fully Convolutional Network}
\acro{CNN}{Convolutional Neural Network}
\acro{GAP}{Global Average Pooling}
\acro{FC}{Fully Connected}
\end{acronym}

\end{document}